\documentclass[%
 reprint,
superscriptaddress,
preprintnumbers,
nofootinbib,
 amsmath,amssymb,
 aps,
]{revtex4-1}

\usepackage{graphicx}
\usepackage{dcolumn}
\usepackage{bm}
\usepackage{hyperref}
\usepackage[mathlines]{lineno}

\usepackage{color}
\usepackage{multirow}

\begin{document}

\preprint{arXiv:2003.02792}

\title{Prospects and requirements of opaque detectors in accelerator neutrino experiments}

\author{Jian Tang}
 \email{tangjian5@mail.sysu.edu.cn}
\author{Sampsa Vihonen}
 \email{sampsa@mail.sysu.edu.cn}
\author{TseChun Wang}
 \email{wangzejun@mail.sysu.edu.cn}
\affiliation{School of Physics, Sun Yat-sen University, Guangzhou 510275, China
}


\begin{abstract}
Opaque detectors are a recently proposed novel detector concept where an opaque scintillator aligned with wavelength-shifting fibers is used to enable the discrimination of electron neutrinos and antineutrinos with a rather low energy threshold. In this work, we investigate the potential effects of the enhanced detection capabilities of the opaque detectors in accelerator neutrino experiments. Focusing on the energy threshold, energy resolution, detection efficiency and background suppression in the analysis of electron-like events, we determine whether using opaque detectors could lead to improvements in the {\sl CP} violation and light sterile neutrino searches in the future accelerator neutrino experiments. We also identify the minimum requirements for the opaque detectors to reach the designated physics goals in the simulated experiments. We find that a 75.6\% fraction of $\delta_{CP}$ values could be reached for {\sl CP} violation discovery by 3$\,\sigma$ confidence level or better when opaque detectors of 120~kton and 130~kton fiducial masses are used together with neutrino beams from J-PARC and MOMENT, respectively, whereas near detectors placed about 250~m from sources are sufficient to exclude the gallium anomaly at 2$\,\sigma$ confidence level.
\end{abstract}


\maketitle

\section{Introduction}
The success of neutrino physics is largely owed to the rapid development of neutrino detection techniques in the recent decades. One of the most established detector technologies to date is the liquid scintillator technique, where neutrino interactions are reconstructed by observing the scintillation light coming from the secondary particles scattering in the detector. Liquid scintillators have facilitated very successful measurements on the neutrino oscillation parameters in reactor neutrino experiments like Daya Bay~\cite{Adey:2018zwh}, Double Chooz~\cite{DoubleChooz:2019qbj} and RENO~\cite{Seo:2015yqp}, solar and geoneutrino detectors Borexino~\cite{Agostini:2019dbs} and KamLAND~\cite{Abe:2008aa}, and most recently in the accelerator neutrino experiment NO$\nu$A~\cite{Adamson:2016xxw}. Over the next decade liquid scintillators will be exploited on the large-scale in experiments like Jingping~\cite{JinpingNeutrinoExperimentgroup:2016nol}, JUNO~\cite{An:2015jdp} and SNO+~\cite{Andringa:2015tza}. Other notable detector techniques for accelerator neutrinos are Liquid Argon Time Projection Chambers (LArTPC) and Water Cherenkov detectors (WC), which have been used in, e.g., ICARUS~\cite{Rubbia:2011ft} and Super-Kamiokande~\cite{Abe:2010hy}.

Most recently, a novel detector concept based on the LiquidO technique has been proposed~\cite{Cabrera:2019rjw}. In this work, we refer to this technology as the opaque detector. Founded on the same detection principle that was used in the famous experiment by Cowan and Reines, opaque detectors are liquid scintillators with short scattering length and intermediate absorption length. In contrast to the conventional model of liquid scintillators, the working principle of the opaque detectors is based on the observation of stochastic light confinement, where the optical photons emitted upon the scintillation revolve around the energy deposition, creating an image of a light ball. The light is collected with an array of wavelength-shifting optical fibers, which run through the detector vessel. The diverging topologies of the light balls in different event categories allow a clear distinction of electron, positron and gamma events at relatively low energies. The LiquidO technique retains the many advantages of an organic scintillator, while it also introduces the imaging capabilities similar to time projection chambers. Most importantly, the relaxation of the transparency requirements grants opaque detectors an ability to load metal-dopants to unprecedented concentrations, as a scalable technology.

Opaque detectors offer a wide array of opportunities in neutrino physics thanks to their relaxed constraints on transparency. Applications for detectors using the LiquidO technique, for example, have been discussed previously in context of neutrinoless double beta decay measurements~\cite{Cabrera:2019rjw}, {\sl CP} violation tests in pion decay-at-rest experiments~\cite{Grassi:2018hew} and testing unitarity of the Pontecorvo-Maki-Nagakawa-Sakata matrix in reactor experiments~\cite{Cabrera:2019xkf}. On the R{\&}D side, the first proof-of-principle experiment on the LiquidO concept has been conducted, and the phenomenon of stochastic light scattering was successfully observed~\cite{Cabrera:2019kxi,Buck:2019tsa}.

In the present work, we consider the possibility of using opaque detectors in accelerator neutrino experiments, including pion and muon decay-in-flight beam options. We first present the expected sensitivities in the standard three-neutrino oscillation framework, reviewing the applications of the opaque detector technique in {\sl CP} violation searches at future accelerator neutrino experiments. More specifically, we discuss what properties would be needed in the detector response and size of the opaque detector to reach desirable sensitivities in neutrino oscillation physics. In this regard, we identify the desired properties for an opaque detector to reach similar sensitivities to what could be achieved in the next-generation Tokai-to-HyperKamiokande (T2HK) experiment~\cite{Abe:2015zbg}. The second topic we choose to discuss is the light sterile neutrino search in the $\Delta m_{41}^2 \simeq$ 1~eV$^2$ regime, which is well motivated by the renowned gallium~\cite{Hampel:1997fc,Abdurashitov:2005tb,Giunti:2010zu} and reactor anomalies~\cite{Mention:2011rk}, and the recently reaffirmed LSND--MiniBooNE anomaly~\cite{Aguilar:2001ty,AguilarArevalo:2007it,Aguilar-Arevalo:2018gpe} (see also Refs.~\cite{Diaz:2019fwt,Gariazzo:2015rra,Boser:2019rta} for the related discussion). Previously studies on the light sterile neutrino search in the future pion and muon-decay-based neutrino sources have been done for e.g. ESSnuSB~\cite{Blennow:2019bvl,Agarwalla:2019zex,Ghosh:2019zvl} and nuSTORM~\cite{Adey:2013pio,Tunnell:2012nu,Adey:2014rfv}, respectively. We compare the sensitivities of T2HK and MOMENT--like beam experiments and opaque detectors with the expected sensitivities of the nuSTORM experiment, which we also provide in this work. We study the sensitivities to the sterile neutrino parameters relevant to the aforementioned neutrino anomalies. Finally, we investigate how using opaque detectors can reduce the impact of the $3+1$ mixing on the {\sl CP} violation sensitivity, which is known to decrease upon the introduction of the new physics~\cite{Choubey:2017ppj,Gandhi:2015xza,deGouvea:2016pom,Dutta:2016glq,Choubey:2018kqq,Choubey:2017cba,Gupta:2018qsv,Chatla:2018sos,Agarwalla:2019zex,Agarwalla_2018,Agarwalla:2016xxa,Ge:2016xya,Ge:2016dlx}.

The simulation studies are based on the General Long-baseline Experiment Simulator (GLoBES)~\cite{Huber:2004ka,Huber:2007ji}. We simulate the performance of a neutrino beam created through pion and muon decay processes while using a large, $\mathcal{O}$(100~kton) opaque detector as a target. As an example of pion and muon decay based neutrino beams, we take the future J-PARC~\cite{Itow:2001ee,Abe:2015zbg} and MOMENT~\cite{Cao:2014bea} facilities assuming 1.3~MW and 15~MW proton beam powers as demonstration, respectively. In this work, we study the effect of the enhanced particle identification in the opaque detector through the energy threshold, energy resolution, detection efficiency and background suppression in electronlike events. As the capabilities of the LiquidO technique are still under active investigation, we treat these parameters as unknowns and simulate the detector performance assuming various possible scenarios. In our simulation, the opaque detector is modeled after the conventional liquid scintillator with enhanced particle identification capabilities and low-energy threshold for electron-like events. The modifications are based on the characteristics that have been reported for the LiquidO technique in Ref.~\cite{Cabrera:2019kxi}.

This article is organized as follows: In Section~\ref{Sec:2}, we briefly describe the key characteristics of an opaque detector and compare its properties with the other notable neutrino detector techniques. We give a brief overview of the neutrino oscillation probabilities in Section~\ref{Sec:Nu_OSC} and detail the implementation of the opaque detector and the neutrino sources in GLoBES in Section~\ref{Sec:3}. We present a discussion on the applications of the opaque detector technique in the {\sl CP} violation and light sterile neutrino searches in Section~\ref{Sec:4}. We summarize our findings in Section~\ref{Sec:5}.

\section{A comparison of opaque detector and other techniques}
\label{Sec:2}

Liquid scintillators have been widely used as low-energy neutrino detectors, and their use in experiments continues to this day. The advantages of using a transparent liquid scintillator include high purity, low energy threshold and detector homogeneity as well as flexible handling and scalability to large volumes. Though effective, transparent liquid scintillators have several requirements that set limitations to their structure and size.

The main challenge involved in using a transparent liquid scintillator is the inability to identify electrons from positrons and gammas at low energies. Adequate charge-identification can be achieved by applying a magnetic field or loading the scintillator with a metal. The main restrictions in loading a scintillator with heavy elements are given by the strict requirements on the solubility, transparency as well as the chemical stability of the scintillator (see Ref.~\cite{Buck:2016vxe} for a review). In practice, the solubility requirement sets an upper limit for the amount of metal that can be loaded in the scintillator, which can vary from 0.1\% through 10\% of the detector mass and depends on the specific application. The difficulties of loading the scintillator beyond those levels arise from dissolving highly polar material into nonpolar liquid, while maintaining optical and chemical stability over the lifetime of the detector.

Abandoning the requirements for optical transparency has immediate advantages regarding the building costs and physics performance of the detector:
\begin{enumerate}
  \item An opaque detector is effectively self-segmented. This could increase the light level, thus helping with the calorimetric precision.
  \item Opaque detectors can have excellent particle identification for electrons and positrons in MeV energies and above~\cite{PedroOchoa}.
  \item Opaque detectors can be expected to scale to larger fiducial masses, thanks to the absence of segmentation.
  \item Opaque detectors can be loaded to high doses of metals. This opens an opportunity to optimize the scintillator composition for various applications.
  \item The opaque detector can be magnetized for further improving the particle identification.
  \item The opaque detector does not need cryogenics.
\end{enumerate}
In summary, neutrino detectors filled with opaque scintillators can provide the imaging capabilities of a time projection chamber and the low energy threshold of a transparent liquid scintillator. With the additional option for magnetization, opaque detectors can be attractive candidate for accelerator-based neutrino experiments.

It should be noted that when the neutrino energies are above the critical value $E_c \sim$ 100~MeV, some of the more energetic electron and positron events lead to electromagnetic showering~\cite{Grassi:2018hew}. Electromagnetic showers are produced when electrons and positrons are created via hard electron-electron scattering or photon bremsstrahlung. This could reduce the $e^- / e^+$ particle identification performance both in efficiency and background rejection. The difficulty of achieving high $e^- / e^+$ discrimination level in electromagnetic showers has been discussed in e.g. Ref.~\cite{Castiglioni:2020tsu}. In opaque scintillators, the loss from electromagnetic showering could be at least partly mitigated with a magnetic field or loading the scintillator with gadolinium.

Opaque detectors may also provide a good alternative to large water Cherenkov detectors. A traditional water Cherenkov detector of $\mathcal{O}$(100~kton) mass scale requires a huge tank and high photo-coverage in light collection. Water Cherenkov detectors receive directional information from the Cherenkov light with a good efficiency and relatively low light yield. This information is used to reconstruct tracks created by muon neutrinos and shower Cherenkov rings by electron neutrinos. Water Cherenkov detectors cannot discriminate positrons from electrons without magnetization, but in quasielastic processes particle identification can be achieved for positrons by doping the water with gadolinium. Gd-doping is also feasible in scintillators, where even higher concentrations could be possible. On the other hand, a typical scintillator detector offers a good light yield, which is transmitted by energy losses as neutrinos interact with detector medium. This is the main advantage of liquid scintillators. Further development on combining the Cherenkov and scintillation signals in a single detector have been made in e.g. the proposed THEIA~\cite{Askins:2019oqj} experiment. As the detector size goes up to 100~kton and beyond, the attenuation of light sets high requirements on the transparency of the detector medium. In the LiquidO detector, this problem is partly mitigated as the light is collected via wavelength-shifting fibers.

One could ask why opaque scintillators should be considered for accelerator neutrino experiments. One could in principle keep the traditional scintillator neutrino techniques based on the success of Daya Bay and NO$\nu$A, or investigate using the novel scintillator recipes to incorporate both scintillation and Cherenkov signals as is done for THEIA. As experiments with longer baselines or more powerful neutrino beams are built, the scalability and durability of opaque detectors make them a competitive choice for future experiments. Further studies on the detector response of opaque scintillators are still needed.

\section{Neutrino Oscillations}
\label{Sec:Nu_OSC}

In this section, we review the neutrino oscillation probabilities relevant to the accelerator neutrino experiments considered in this work. We discuss the potential applications in the {\sl CP} violation searches in subsection~\ref{Sec:CPV_theory} and in constraining active-sterile neutrino parameters in the $\Delta m_{41}^2 \simeq$ 1~eV$^2$ regime in subsection~\ref{Sec:Sterile_nu}.

In the standard three-neutrino paradigm, neutrino oscillations are governed by the so-called Pontecorvo-Maki-Nakagawa-Sakata (PMNS) matrix~\cite{Pontecorvo:1957cp,Pontecorvo:1957qd,Maki:1960ut,Maki:1962mu,Pontecorvo:1967fh}, which describes the mixing with three rotation angles $\theta_{12}$, $\theta_{13}$, $\theta_{23}$ and the Dirac {\sl CP} phase $\delta_{CP}$:

\begin{widetext}
\begin{equation}\label{eq:PMNS}
\begin{split}
U_\text{PMNS} & =  R_{3\times3}(\theta_{23}, 0)\times R_{3\times3}(\theta_{13}, \delta_{CP})\times R_{3\times3}(\theta_{12}, 0)\\
& =
\left(
\begin{array}{ccc}
1 & 0 & 0\\
0 & c_{23} & s_{23} \\
0 & -s_{23} & c_{23}
\end{array}\right) 
\left(
\begin{array}{ccc}
c_{13} & 0 & s_{13}\mathrm{e}^{-i\delta_\mathrm{CP}}\\
0 & 1 & 0 \\
s_{13}\mathrm{e}^{i\delta_\mathrm{CP}} & 0 & c_{13}
\end{array}\right) 
\left(
\begin{array}{ccc}
c_{12} & s_{12} & 0\\
-s_{12} & c_{12} & 0 \\
0 & 0 & 1
\end{array}\right),
\end{split}
\end{equation}
\end{widetext}
where $R_{3\times 3}(\theta_{ij},\delta_{ij})$ is a $3 \times 3$ rotation matrix with the rotation angle $\theta_{ij}$ and the {\sl CP} phase $\delta_{ij}$ ($i$,~$j =$ 1, 2, 3 and $\delta_{13} \equiv \delta_{CP}$), while $s_{ij}$ and $c_{ij}$ denote the functions $\sin \theta_{ij}$ and $\cos \theta_{ij}$, respectively. 

The probability for the oscillation $\nu_\alpha \rightarrow \nu_\beta$ ($\alpha$,~$\beta = e$,~$\mu$,~$\tau$) to take place in vacuum is given by
\begin{equation}\label{eq:P}
\begin{array}{l}
P(\nu_\alpha\rightarrow\nu_\beta) = \\
\delta_{\alpha\beta}-4\sum_{k>j}\mathrm{Re}[U^*_{\alpha k}U_{\beta k}U_{\alpha j}U^*_{\beta j}]\sin^2\left(\frac{\Delta m_{kj}^2 L}{4E}\right)\\
\hspace{0.7cm}
+2\sum_{k>j}\mathrm{Im}[U^*_{\alpha k}U_{\beta k}U_{\alpha j}U^*_{\beta j}]\sin\left(\frac{\Delta m_{kj}^2 L}{2E}\right),
\end{array}
\end{equation}
where $U_{ij}$ are the matrix elements of Eq.~(\ref{eq:PMNS}), $\Delta m_{ij} \equiv m_i^2 - m_j^2$ are the so-called mass-squared differences, whereas $E$ is the energy of the neutrino and $L$ is the distance it has traveled.

In the present work, we take the values for the neutrino oscillation parameters from the current best-fits that were obtained from the global neutrino oscillation data by the NuFit collaboration~\cite{NuFit:4-1}. The central values and standard deviations used in our simulations are provided in Table~\ref{bounds}. Our analysis is based on the $\chi^2$ method described in Refs.~\cite{Huber:2004ka,Huber:2007ji}.

\begin{table}[!t]
\caption{\label{bounds} The best-fit values and 1$\,\sigma$ confidence level uncertainties for the neutrino oscillation parameters reported by the NuFit group~\cite{NuFit:4-1}. The values are shown for both normal ordering (NO) and inverted ordering (IO), where $\Delta m_{3\ell}^2$ corresponds to $\Delta m_{31}^2$ (NO) and  $\Delta m_{32}^2$ (IO), respectively.}
\begin{center}
\resizebox{\linewidth}{!}{%
\begin{tabular}{ccccc}\hline\hline
Parameter & Central value $\pm$ 1\,$\sigma$ (NO) \,\,\, & Central value $\pm$ 1\,$\sigma$ (IO) \\ \hline
\rule{0pt}{3ex}$\theta_{12}$ ($^\circ$) & 33.820 $\pm$ 0.780 & 33.820 $\pm$ 0.780 \\ 
\rule{0pt}{3ex}$\theta_{13}$ ($^\circ$) & 8.600 $\pm$ 0.130 & 8.640 $\pm$ 0.130 \\ 
\rule{0pt}{3ex}$\theta_{23}$ ($^\circ$) & 48.600 $\pm$ 1.400 & 48.800 $\pm$ 1.200 \\ 
\rule{0pt}{3ex}$\delta_{CP}$ ($^\circ$) & 221.000 $\pm$ 39.000 & 282.000 $\pm$ 25.000 \\ 
\rule{0pt}{3ex}$\Delta m_{21}^2$ (10$^{-5}$ eV$^2$) & 7.390 $\pm$ 0.210 & 7.390 $\pm$ 0.210 \\ 
\rule{0pt}{3ex}$\Delta m_{3l}^2$ (10$^{-3}$ eV$^2$) & 2.528 $\pm$ 0.031 & -2.510 $\pm$ 0.031 \\ \hline\hline
\end{tabular}}
\end{center}
\end{table}

\subsection{Searching {\sl CP} violation with $\nu_\mu$ source}\label{Sec:CPV_theory}

One of the main goals for the next-generation of neutrino experiments is to search for the {\sl CP} violation arising from $\delta_{CP}$ and to determine its size. In accelerator-based neutrino experiments, {\sl CP} violation can be tested by comparing probabilities for $\nu_\mu \rightarrow \nu_e$ and $\bar{\nu}_\mu \rightarrow \bar{\nu}_e$ (both pion and muon decay based beams) and for $\nu_e \rightarrow \nu_\mu$ and $\bar{\nu}_e \rightarrow \bar{\nu}_\mu$ (only muon decay based beams). 

Taking $\xi \equiv \frac{\Delta m^2_{21}}{|\Delta m_{31}^2|}$ to be small and $s^2_{13} = \sin^2 \theta_{13} \sim \mathcal{O}(\xi)$, this probability can be expressed as
\begin{equation}\label{Eq:CPV1}
P(\nu_\mu\rightarrow\nu_e)= P_1 + P_{\frac{3}{2}} + \mathcal{O}(\xi^2),
\end{equation}
where the first two terms are given by 
\begin{equation}
P_1=\frac{4}{(1-r_A)^2}s^2_{23}s^2_{13}\sin^2\left(\frac{(1-r_A)\Delta_{31}}{2}\right),
\end{equation}

\begin{equation}\label{Eq:CPV2}
\begin{array}{l}
P_{\frac{3}{2}}=8J_r\frac{\xi}{r_A(1-r_A)}\cos\left(\delta_\mathrm{CP}+\frac{\Delta_{31}}{2}\right)\sin\left(\frac{r_A\Delta_{31}}{2}\right)
\\
\hspace{0.7cm}\times\sin\left(\frac{(1-r_A)\Delta_{31}}{2}\right),
\end{array}
\end{equation}
where $J_r = c_{12} \, s_{12} \, c_{23} \, s_{23} \, s_{13}$, $r_A = a_e / \Delta m_{31}^2$ and $\Delta_{31} \equiv \Delta m_{31}^2 L / 2E$. Here $a_e \equiv 2\sqrt{2} \, E \, G_F \, N_e$, where $G_F$ is the Fermi coupling constant and $N_e$ is the electron number density in the Earth. 

As one can see from Eqs.~(\ref{Eq:CPV1}--\ref{Eq:CPV2}), the Dirac {\sl CP} phase $\delta_\mathrm{CP}$ appears in the order of $\mathcal{O}(\xi^{3/2})$ inside a cosine function together with $\Delta_{31}/2$. This function reaches its first maximum at $\Delta_{31}/2 = \pi/2$ and second maximum at $3\pi/2$, respectively. Experiments covering either or both of these oscillation maxima can therefore be expected to be sensitive to $\delta_{CP}$.

\subsection{Probability with the active-sterile mixing}\label{Sec:Sterile_nu}

Another well-motivated question to be investigated at the future experiments is the LSND-MiniBooNE anomaly and the effect of a light sterile neutrino of $\mathcal{O}$(1~eV) mass in the neutrino oscillation probabilities. This question is also supported by the gallium and reactor anomalies. When a light sterile neutrino is considered, the PMNS matrix becomes a submatrix of a 4$\times$4 mixing matrix, which is given by
\begin{equation}\label{Eq:U_3+1}
\begin{array}{l}
U_{3+1}=R_{4\times4}(\theta_{34},\delta_{34})R_{4\times4}(\theta_{24},\delta_{24})R_{4\times4}(\theta_{14}, 0)\\ \hspace{1.25cm}\times R_{4\times4}(\theta_{23},0)R_{4\times4}(\theta_{13},\delta_\mathrm{CP})R_{4\times4}(\theta_{12},0),
\end{array}
\end{equation}
with three new mixing angles $\theta_{14}$, $\theta_{24}$, and $\theta_{34}$, and two new phases $\delta_{24}$ and $\delta_{34}$. To calculate oscillation probabilities, we also need to include a new mass-squared splitting $\Delta m_{41}^2 \sim$ 1~eV$^2$.

In the following, we apply the so-called short-baseline approximation $\Delta_{41} \equiv \Delta m_{41}^2 L / 2E \gg \Delta_{31}$ on Eq.~(\ref{eq:P}). For measuring active-sterile mixing parameters in a near detector, the probabilities are given by~\cite{Meloni:2010zr}
\begin{equation}\label{eq:Pmumu_ND}
\begin{array}{l}
\hspace{0.5cm}P(\nu_\mu\rightarrow\nu_\mu)=1-4\cos^2\theta_{14}\sin^2\theta_{24}\\
\hspace{2.7cm}
\times\left(1-\cos^2\theta_{14}\sin^2\theta_{24}\right)\sin^2(\Delta m_{41}^2L/4E)
\end{array}
\end{equation}
and
\begin{equation}\label{eq:Pmue_ND}
\begin{array}{l}
\hspace{-1.cm}P(\nu_\mu\rightarrow\nu_e)=\sin^22\theta_{14}\cos^22\theta_{24}\\
\hspace{1.2cm}
\times\sin^2\theta_{34}\sin^2(\Delta m_{41}^2L/4E).
\end{array}
\end{equation}

Near detectors can be optimized for the measurement of the active-sterile mixing parameters. In a beam experiment with $E \sim \mathcal{O}$(100~MeV), baseline lengths of $L \sim \mathcal{O}$(100~m) are the most sensitive to the $\Delta m_{41}^2 \sim$ 1~eV$^2$ scale.

In the far detectors, the oscillations are dominated by $\Delta m_{31}^2$ and $\Delta m_{41}^2 \sim \mathcal{O}$(1~eV$^2$) are averaged out. After taking $s^2_{14}\sim s^2_{24}\sim s^2_{34}\sim s^2_{13}\sim \mathcal{O}(\xi^2)$ to be small, the oscillation probabilities for $\nu_\mu \rightarrow \nu_\mu$ and $\nu_\mu \rightarrow \nu_e$ can be approximated as~\cite{Meloni:2010zr}
\begin{equation}\label{eq:Pmumu_FD}
\begin{array}{l}
P(\nu_\mu\rightarrow\nu_\mu)\approx \sin^2\left(\Delta_{31}\right)\left(1-2s^2_{24}\right)+8\hat{s}^2_{23}\sin^2\left(\Delta_{31}\right)\\
\hspace{1.5cm}+2s_{24}s_{34}\cos(\delta_{34}+\delta_{24})\Delta_n\sin\left(2\Delta_{31}\right)\\
\hspace{1.5cm}-s_{13}^2\Delta_{31}\cos\left(\Delta_{31}\right)\times\left(\left(\Delta_{31}-\Delta_e\right)\Delta_e\sin(\Delta_{31})
\right.\\ 
\hspace{1.5cm}
\left. -\Delta_{31}\sin(\Delta_{31}-\Delta_e)\sin(\Delta_e)\right)/(\Delta_{31}-\Delta_e)^2
\end{array}
\end{equation}
and
\begin{equation}\label{eq:Pmue_FD}
\begin{array}{l}
\hspace{-0.34cm}
P(\nu_\mu\rightarrow\nu_e)\approx 1-\left\{\cos^2(\Delta_{31})(1-2s^2_{24})+2s^2_{13}\Delta_{31}\cos(\Delta_{31})\right. \\ \hspace{1.9cm}
\times \frac{\Delta_{31}\sin(\Delta_{31}-\Delta_e)\sin(\Delta_e)}{(\Delta_{31}-\Delta_e)^2}+\sin^2(\Delta_{31})(1-s_{24}^2) \\ \hspace{1.9cm} 
-s_{13}^2\Delta_{31}
\sin\Delta_{31}\frac{\Delta_{31}\left[\sin(\Delta_{31}-\Delta_e)+\sin(\Delta_e)\right]}{(\Delta_{31}-\Delta_e)^2}
\\ \hspace{1.9cm} \left. +\frac{1}{2}s_{24}^2\left(3+\cos(2\Delta_{31})\right)\right\},
\end{array}
\end{equation}
where $\hat{s}_{23}^2 = \sin^2 \theta_{23} - 1/\sqrt{2}$, $\Delta_{e(n)}\equiv a_{e(n)}L/4E$ and $a_{n}\equiv 2\sqrt{2} \, E \, G_F \, N_{n}$, whereby $N_n$ is the neutron number density in the Earth.

As was pointed out in Ref.~\cite{Meloni:2010zr}, a resonance takes place in Eqs.~(\ref{eq:Pmumu_FD}) and (\ref{eq:Pmue_FD}) when $\Delta m^2_{41}\sim \Delta m_{31}^2$. In principle, far detectors with $L\sim\mathcal{O}$(100~km) are suitable for measuring the sterile mixing angles. However, it is unavoidable that the new angles $\theta_{24}$ and $\theta_{34}$ will hinder their measurement~\cite{Choubey:2017ppj}. The presence of the sterile phases could also reduce the sensitivity to discover {\sl CP} violation~\cite{Gandhi:2015xza,deGouvea:2016pom,Dutta:2016glq}.

\section{Implementation in GLoBES}
\label{Sec:3}

In this work, we study the prospects and requirements of sending a neutrino/antineutrino beam from an accelerator facility to a detector employing the LiquidO technique. We focus on the energy range relevant to medium and long-baseline experiments, $E_{\nu,\bar{\nu}} \sim \mathcal{O}$(100~MeV), where neutrinos and antineutrinos interact with the scintillator predominantly via quasi-elastic scattering. We obtain our results using the General Long-baseline Experiment Simulator (GLoBES)~\cite{Huber:2004ka, Huber:2007ji}. 

We consider two different neutrino beams as the source, one based on the pion decay and the other on the muon decay. For the pion-decay-based neutrino beam, we simulate the JHF beam from the J-PARC facility~\cite{Itow:2001ee}, which will operate at 1.3~MW proton beam power after its next upgrade~\cite{Abe:2015zbg}. The JHF beam is currently proposed to be used in the T2HK experiment, which includes a water Cherenkov detector in its configuration. As for the muon decay based neutrino beam, we consider the MOMENT facility~\cite{Cao:2014bea} with its 15~MW proton beam power. The properties of the neutrino beams are summarized in Table~\ref{NuSources}. We assume these beam configurations in our simulations throughout this work, unless otherwise stated.

\begin{table}[!t]
\caption{\label{NuSources}Details for the experimental parameters in the J-PARC and MOMENT beam facilities in our simulation.}
\begin{center}
\begin{tabular}{ccc}
\hline\hline
\rule{0pt}{3ex}Facility & J-PARC & MOMENT  \\ \hline
\rule{0pt}{3ex}Source location & Japan & China  \\ 
\rule{0pt}{3ex}Production mechanism & pion decay & muon decay \\ 
\rule{0pt}{3ex}Expected beam power & 1.3~MW & 15~MW \\ 
\rule{0pt}{3ex}Expected energy range & 0...1.2~GeV & 0...800~MeV \\ 
\rule{0pt}{3ex}Baseline length & 295~km & 150~km \\ 
\rule{0pt}{3ex}Operational time & 2.5+7.5 years & 5+5 years \\ \hline\hline
\end{tabular}
\end{center}
\end{table}

\begin{table}[!t]
\caption{\label{NuDetector} Properties of the opaque detector assumed in the simulation. Unless otherwise stated, the detector has 187~kton fiducial mass and an energy range divided into 20 equisized bins. Here $\sigma_e$ corresponds to the Gaussian width of the energy resolution function.}
\begin{center}
\begin{tabular}{ccc}
\hline\hline
\rule{0pt}{3ex}Event type & $e$-like & $\mu$-like \\ \hline
\rule{0pt}{3ex}Detection efficiencies & 50\% & 80\% \\
\rule{0pt}{3ex}Energy threshold & 10~MeV & 100~MeV \\
\rule{0pt}{3ex}$\sigma_e = \beta \, \sqrt{E}$ & $\beta =$ 10\% & $\beta =$ 5\% \\ \hline\hline
\end{tabular}
\end{center}
\end{table}

\begin{table*}[!t]
\caption{\label{NuConfig} Summary of the signal and background channels that have been assumed for the opaque detector as far detector. The background composes of the beam-related impurities and neutral currents in T2HK--like experiments, whereas MOMENT--like experiments get background from charge-misidentification, neutral currents and atmospheric neutrinos. The last two columns show the total number of signal and background events in a 100~kton opaque detector. The opaque detector is assumed to have 50\% efficiency for electronlike events and 80\% for muonlike events and 10 years of beam exposure.}
\begin{center}
\begin{tabular}{c@{\hskip 8ex}c@{\hskip 8ex}c@{\hskip 8ex}c@{\hskip 8ex}c}
\hline\hline
\rule{0pt}{3ex}Neutrino source & Signal & Background & Signal events & Background events \\ \hline
\rule{0pt}{3ex}\multirow{4}{*}{J-PARC} & $\nu_\mu \rightarrow \nu_e$ & $\nu_\mu \rightarrow \nu_\mu$ CC, $\nu_e \rightarrow \nu_e$ CC, $\nu_\mu$ NC & 523 & 25\\
\rule{0pt}{3ex} & $\nu_\mu \rightarrow \nu_\mu$ & $\nu_\mu$ NC & 3721 & 14\\
\rule{0pt}{3ex} & $\bar{\nu}_\mu \rightarrow \bar{\nu}_e$ & $\bar{\nu}_\mu \rightarrow \bar{\nu}_\mu$ CC, $\bar{\nu}_\mu$ NC & 285 & 40\\
\rule{0pt}{3ex} & $\bar{\nu}_\mu \rightarrow \bar{\nu}_\mu$ & $\bar{\nu}_\mu$ NC & 3265 & 12\\ \hline
\rule{0pt}{3ex}\multirow{8}{*}{MOMENT} & $\bar{\nu}_\mu \rightarrow \bar{\nu}_e$ & $\nu_e \rightarrow \nu_e$ CC, $\bar{\nu}_\mu$ NC, $\nu_e$ NC, atmospheric & 36 & 41\\
\rule{0pt}{3ex} & $\bar{\nu}_\mu \rightarrow \bar{\nu}_\mu$ & $\nu_e \rightarrow \nu_\mu$ CC, $\bar{\nu}_\mu$ NC, $\nu_e$ NC, atmospheric & 576 & 24\\
\rule{0pt}{3ex} & $\nu_e \rightarrow \nu_e$ & $\bar{\nu}_\mu \rightarrow \bar{\nu}_e$ CC,  $\bar{\nu}_\mu$ NC, $\nu_e$ NC, atmospheric & 3093 & 23\\
\rule{0pt}{3ex} & $\nu_e \rightarrow \nu_\mu$ & $\bar{\nu}_\mu \rightarrow \bar{\nu}_\mu$ CC, $\bar{\nu}_\mu$ NC, $\nu_e$ NC, atmospheric & 187 & 25\\
\rule{0pt}{3ex} & $\bar{\nu}_e \rightarrow \bar{\nu}_e$ & $\nu_\mu \rightarrow \nu_e$ CC, $\bar{\nu}_e$ NC, $\nu_\mu$ NC, atmospheric & 480 & 16\\
\rule{0pt}{3ex} & $\bar{\nu}_e \rightarrow \bar{\nu}_\mu$ & $\nu_\mu \rightarrow \nu_\mu$ CC, $\bar{\nu}_e$ NC, $\nu_\mu$ NC, atmospheric & 33 & 18\\
\rule{0pt}{3ex} & $\nu_\mu \rightarrow \nu_\mu$ & $\bar{\nu}_e \rightarrow \bar{\nu}_\mu$ CC, $\bar{\nu}_e$ NC, $\nu_\mu$ NC, atmospheric & 759 & 15\\
\rule{0pt}{3ex} & $\nu_\mu \rightarrow \nu_e$ & $\bar{\nu}_e \rightarrow \bar{\nu}_e$ CC, $\bar{\nu}_e$ NC, $\nu_\mu$ NC, atmospheric & 93 & 18\\ \hline\hline
\end{tabular}
\end{center}
\end{table*}

Regarding the implementation of the far detector, we assume a large opaque detector with the fiducial masses of 50~kton...500~kton. In the case of the J-PARC beam facility, we assume the detector to be placed at 295~km with 2.5$^\circ$ off-axis angle, which corresponds to the configuration where the JHF beam reaches the first oscillation maximum. In the case of MOMENT, the opaque detector is taken to be 150~km from the beam facility, facing the neutrino/antineutrino beam on-axis. We assume both configurations to operate 10 years in total, which is divided in 1:3 and 1:1 ratios in order to provide similar rates for neutrino and antineutrino events from J-PARC and MOMENT beams, respectively.

As the detector response of the opaque detectors is still under active investigation, we simulate the opaque detector following the example set by the NO$\nu$A far detector, which is based on a 14~kton segmented liquid scintillator. To account for the improvements that can be introduced using an opaque detector, we assume a 80\% detection efficiency for muonlike events and 50\% efficiency for electronlike events~\footnote{It has been suggested that in the LiquidO detector the efficiencies can be as high as 100\% for both $e$ and $\mu$-like events~\cite{Anatael}. In our simulations, we maintain the conservative choices of 50\% and 80\%, respectively.}. The energy of each recorded event is smeared according to their incident energy $E$ and a Gaussian energy resolution function, which is defined by the width function $\sigma_e = \beta \, \sqrt{E}$. In a liquid scintillator, the energy resolution is generally given as $\beta =$ 10\% for electronlike events and 5\% for muonlike events, respectively. The details concerning the detector simulation are summarized in Table~\ref{NuDetector}.

Thanks to their improved imaging capabilities, opaque detectors can help to constrain the number of background events in both pion- and muon-decay-based neutrino experiments. In Table~\ref{NuConfig}, we list the signal and background components that are used in the pion-decay and muon-decay-based neutrino beam configurations when the far detector adopts the opaque detection principles. In the JHF beam generated in J-PARC, the background to the $\nu_\mu$ and $\bar{\nu}_\mu$ beams consists of the intrinsic beam background, flavor misidentification and neutral current events, of which the beam background is irreducible. MOMENT on the other hand produces neutrinos and antineutrinos in $\nu_e \bar{\nu}_\mu$ and $\bar{\nu}_e \nu_\mu$ pairs, which have negligible beam-related backgrounds. Instead, it accumulates background events from charge misidentifications, atmospheric neutrino backgrounds and neutral current events. We consider the atmospheric neutrino background component to be irreducible. The difference between the events in $\nu_e \rightarrow \nu_e$ and $\bar{\nu}_e \rightarrow \bar{\nu}_e$ arises from the convoluted effect of the differences in the neutrino fluxes, cross sections and oscillations.

We adopt the following treatment of the background events in our baseline simulation. The efficiencies and the composition of the beam-related backgrounds in J-PARC are obtained from Refs.~\cite{Ambats:2004js,Yang_2004}. We assume the J-PARC configuration to acquire 88\% suppression rate for the beam-related backgrounds, 0.04\% acceptance for flavour misidentifications and (0.37\%) 0.15\% acceptance for ($\bar{\nu}_e$) $\nu_e$, $\nu_\mu$ and neutral currents. MOMENT gains backgrounds from charge misidentifications and neutral currents with acceptance rates of 0.3\% and 0.25\%, respectively. For more information about the background composition in MOMENT, see Refs.~\cite{Cao:2014bea,Blennow:2015cmn,Tang:2017khg,Tang:2019wsv}.

The number of signal events corresponding to $\nu_{\alpha} \rightarrow \nu_{\beta}$ ($\alpha$, $\beta = e$, $\mu$, $\tau$) oscillations in our simulations can be calculated as 
\begin{widetext}
\begin{equation}
N_i = T \, N_\text{nucl} \, \epsilon \int_{E_\text{min}}^{E_\text{max}} \int_{E_\text{min}^{'}}^{E_\text{max}^{'}} dE \, d{E'} \, \phi(E) \, \sigma(E) \, R(E, E') \, P_{\nu_{\alpha} \rightarrow \nu_{\beta}}(E),
\label{Events}
\end{equation}
\end{widetext}
where $i$ is the number of the energy bin, $T$ is the total runtime of the experiment and $N_\text{nucl}$ is the number of target nucleons in the detector. The detection efficiency for the final state neutrino $\nu_\beta$ is denoted with $\epsilon$. The integration is done over the true and reconstructed energies $E$ and $E'$, and the neutrino/antineutrino fluxes, cross sections and energy resolution are provided in $\phi(E)$, $\sigma(E)$ and $R(E,E')$, respectively. The oscillation probability for $\nu_{\alpha} \rightarrow \nu_{\beta}$ is included in $P_{\nu_{\alpha} \rightarrow \nu_{\beta}}(E)$.

To quantify the background reduction in the opaque detector, we calculate the signal-to-background ratio for both neutrino sources as the ratio of signal and background events. When the beam is created in the J-PARC facility, the ratio is determined from signal and background events from the $\nu_\mu \rightarrow \nu_e$ channel, whereas for the MOMENT beam it is calculated from the events of the $\nu_e \rightarrow \nu_\mu$ channel. We fix the number of signal events in the far detector, whilst the background is suppressed to reach the desired signal-to-background ratio. The background suppression is applied to all background components except for the irreducible part. In a 100~kton opaque detector, for example, we fix the number of signal events at 523 and 187 for the beams from J-PARC and MOMENT, respectively, while assuming 50\% efficiency for $e$-like events. In detectors of other sizes the signal events scale according to the fiducial mass.

We assume 2.5\% and 5\% systematic uncertainties on the $e$-like and $\mu$-like signal events, respectively, and 5\% systematic uncertainties on all background events when the MOMENT beam is in use. As far as J-PARC is concerned, we take the systematic uncertainties to be 5\% for all signal and background events. The prior function takes the values of the standard neutrino oscillation parameters from Table~\ref{bounds}.

The improved particle identification capabilities and the low energy threshold of the opaque detector technique lead into enhanced statistics in the low energy spectrum. The absence of segmentation could also allow larger fiducial masses compared to transparent liquid scintillators. In our simulation, we study the effects of detection efficiency and energy threshold on the $e$-like events as well as background suppression and fiducial mass on the physics prospects of the considered experiments. Our aim is to find the minimum requirements for the opaque detector technology to reach the sensitivity targets. We also check the effect of energy resolution on the results. The efficiencies are taken into account through parameter $\epsilon$ in Eq.~(\ref{Events}).

\section{Physics prospects and detector requirements}
\label{Sec:4}

In this section, we discuss the prospects and requirements of using a large-scale opaque detector in establishing {\sl CP} violation in accelerator neutrino experiments, and in constraining the active-sterile mixing parameters in the 3+1 model by using a near detector containing opaque scintillator. We present the sensitivities for the {\sl CP} violation discovery using beams from the MOMENT and J-PARC facilities and find the exclusion limits for the parameters describing the active-sterile neutrino mixing. We obtain the minimum requirements for the relevant detector parameters to reach the expected {\sl CP} violation sensitivities of T2HK and to exclude the parameter values currently allowed by the gallium and reactor neutrino anomalies within 2$\,\sigma$ confidence level.

\subsection{Sensitivity to CPV discovery}
\label{Sec:4.1}

\begin{figure*}
\begin{center}
\includegraphics[width=\textwidth]{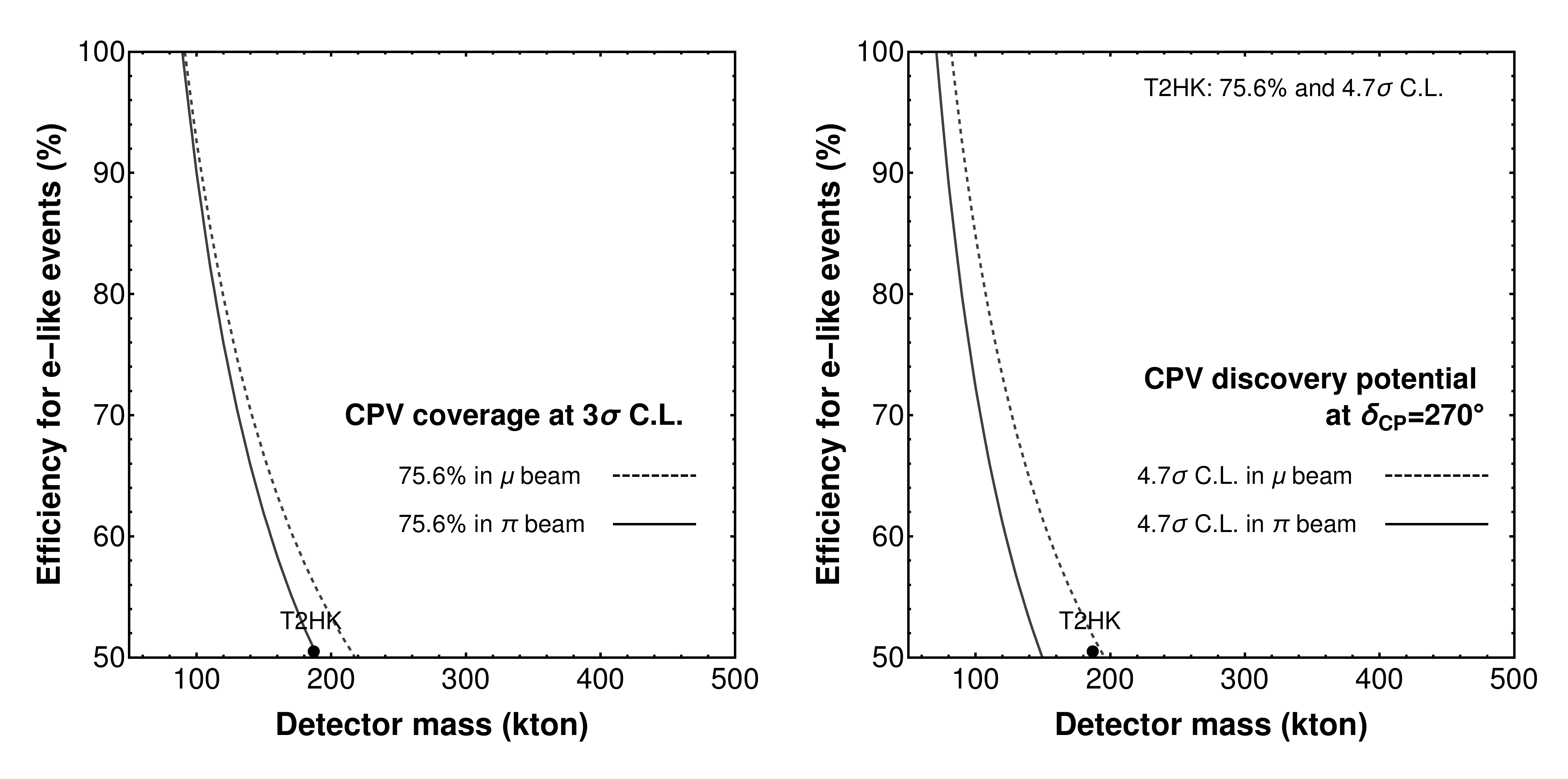}
\end{center}
\caption{\label{CPVplot_e-eff} CP discovery potential in an accelerator neutrino experiment using muon- and pion-decay beams and an opaque detector. Left: the detector masses and efficiencies for which the {\sl CP}-conserving values $\delta_{CP} =$ 0$^\circ$, 180$^\circ$ can be ruled out at 3$\,\sigma$ confidence level (C.L.) or better for 75.6\% of the $\delta_{CP}$ values. Right: the masses and efficiencies required to reach 4.7$\,\sigma$~C.L. statistical significance for {\sl CP} violation discovery when the true value is $\delta_{CP} =$ 270$^\circ$. The approximate position of T2HK in this map is indicated with a black dot.}
\end{figure*}

One of the most important missions of the next-generation accelerator neutrino experiments is to establish {\sl CP} violation in the leptonic sector and measure the value of the Dirac {\sl CP} phase $\delta_{CP}$. The typical challenge in measuring $\delta_{CP}$ arises from the need to collect sufficient number of events from both neutrino and antineutrino channels. In a pion-decay-based neutrino beam experiment, the sensitivity to {\sl CP} violation discovery is mainly acquired from $\nu_\mu \rightarrow \nu_e$ and $\bar{\nu}_\mu \rightarrow \bar{\nu}_e$ channels, which place importance on the reconstruction of electron and positron events and on the background suppression. In this subsection, we investigate whether the advanced particle identification capabilities in the opaque detector could improve the sensitivity to {\sl CP} violation.

In this subsection, we aim to determine the minimum requirements for the detector parameters to reach the desired sensitivity in {\sl CP} violation discovery using muon- and pion-decay-based neutrino beams from J-PARC and MOMENT. To provide a meaningful comparison and sensitivity target for our study, we calculate the expected sensitivities for the T2HK experiment assuming a single 187~kton water Cherenkov detector. We use simulation files provided on the GLoBES website~\cite{GLoBES} (see also Refs.~\cite{Huber:2002mx,Itow:2001ee,Ishitsuka:2005qi}). The beam power, detector mass and running times are updated to match the current state of the T2HK proposal, featuring 1.3~MW beam power, 187~kton WC detector and 2.5 + 7.5 years of running time. The energy threshold is set at 0.4~GeV. With this setup, our results show that {\sl CP} violation discovery can be reached in T2HK at 3$\,\sigma$~C.L. or higher significance for 75.6\% of the theoretically allowed $\delta_{CP}$ values. In the same time, a discovery can be reached at 4.7$\,\sigma$~C.L. significance if the true value is $\delta_{CP} =$ 270$^\circ$.

\begin{figure*}
\begin{center}
\includegraphics[width=\linewidth]{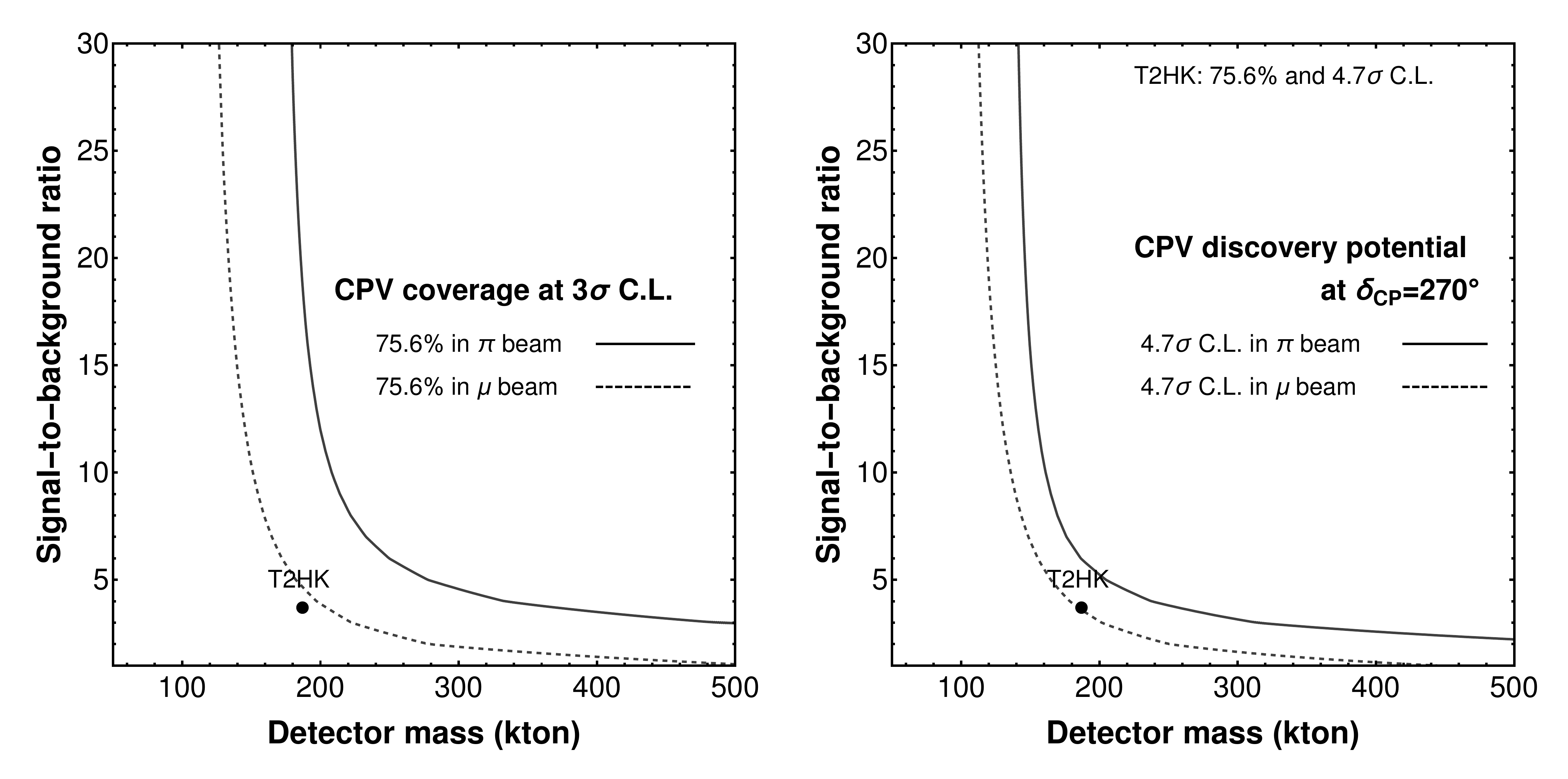}
\end{center}
\caption{\label{CPVplot_sbratio} The same as in Fig.~\ref{CPVplot_e-eff}, but the CP fractions and statistical significances are presented as a function of detector mass and signal-to-background ratio in the configuration. The number of signal events in $\nu_\mu \rightarrow \nu_e$ and $\nu_e \rightarrow \nu_\mu$ are are fixed at 523 and 187 per 100~kton in pion- and muon-decay beams, respectively. The results are shown for signal-to-background ratios 1--30. The approximate position of T2HK in this map is indicated with a black dot.}
\end{figure*}

We have calculated the CPV discovery potential simulating an opaque detector of various sizes and properties with muon-decay- and pion-decay-based based beam setups. The $\chi^2$ fitting has been done by marginalizing over all oscillation parameters except for $\delta_{CP}$. We selected the detector parameter values that allow to establish {\sl CP} violation for 75.6\% of $\delta_{CP}$ and at 4.7$\,\sigma$~C.L. significance for $\delta_{CP} =$ 270$^\circ$ in the considered experiment setup. The results are presented in Figures~\ref{CPVplot_e-eff} and~\ref{CPVplot_sbratio}.

In Figure~\ref{CPVplot_e-eff}, we present the minimum values for the detector efficiencies as function of the fiducial mass of the detector. The efficiencies are imposed on the electronlike events. In both panels, the detector efficiencies and fiducial masses required to reach the {\sl CP} violation sensitivity in the pion-decay-based and muon-decay-based beam setups are indicated by solid and dashed curves, respectively. The fiducial mass and efficiency corresponding to the T2HK configuration, that is 187~kton and 50.498\%, are illustrated with the black dot in this mapping. Figure~\ref{CPVplot_sbratio} on the other hand shows the requirements for the background suppression to reach the target sensitivity. The result is expressed in terms of the signal-to-background ratio and it is presented as function of detector mass. We estimated the signal-to-background ratio to be approximately 3.70 in the $\nu_\mu \rightarrow \nu_e$ channel in the simulated T2HK setup.

One notable advantage of the opaque scintillators with respect to the transparent ones is the low energy threshold. We examined whether the energy threshold has any relevance in the {\sl CP} violation search. To do this, the CP fraction and statistical significance of the J-PARC and MOMENT configurations were re-calculated with inflated energy threshold values, varying the corresponding detector parameter between 10~MeV and 100~MeV. The energy resolution function was also varied, assuming different values for the Gaussian width $\sigma_e$. We considered energy resolutions between 5\% and 20\%.

Basing on our simulation results, we find both the improved detector efficiency and background suppression to influence the prospects for {\sl CP} violation, whereas the low-energy threshold leads to no significant improvement. The effect of energy resolution has similarly little relevance on the physics reach. The minimum requirement for the detector efficiency in electronlike events relaxes as larger detector masses are considered. The correlation is not linear, though, and the detector efficiency has more importance in the {\sl CP} violation discovery when the efficiency is less than 75\%. We observe similar behaviour in the role of background suppression, as the requirement on the signal-to-background ratio relaxes faster for lower fiducial masses, and it becomes more relevant when its value is below 10.

An interesting question about the opaque detector technology is whether it could be useful in the T2HK experiment itself. If the purified water of the water Cherenkov detector were to be replaced with an opaque medium equivalent to 187~kton fiducial mass, the corresponding CPV coverage at 3$\,\sigma$ C.L. and significance at $\delta_{CP} =$ 270$^\circ$ would be 74.3\% and 4.9$\,\sigma$ C.L. with 50\% efficiency for $e$-like events, respectively. Increasing the efficiency to 70\% would raise the {\sl CP} violation sensitivity to 77.1\% coverage at 3$\,\sigma$ C.L. and 5.3$\,\sigma$ C.L. significance at $\delta_{CP} =$ 270$^\circ$. Replacing the water with an equivalent mass of opaque scintillator could therefore improve the prospects for {\sl CP} violation discovery in T2HK, though the gain would be modest. One would therefore have to consider larger fiducial masses for the opaque detector to significantly improve the sensitivity to {\sl CP} violation.

\begin{figure*}
\begin{center}
\includegraphics[width=0.9\linewidth]{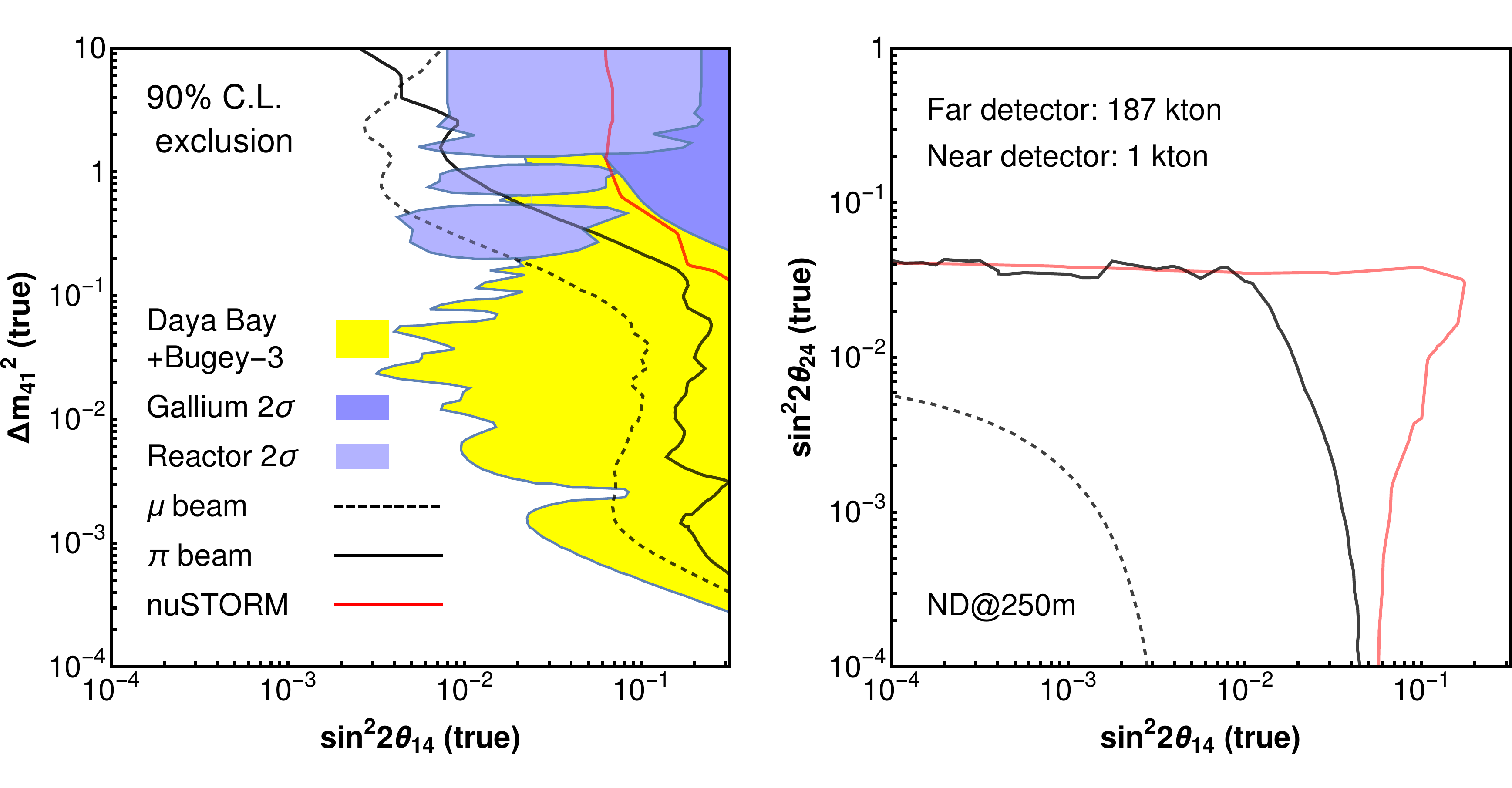}
\end{center}
\caption{\label{SterilePlot} The exclusion limits for $\sin^2 2\theta_{14}$ and $\Delta m_{41}^2$ (left) and for $\sin^2 2\theta_{14}$ and $\sin^2 2\theta_{24}$ (right) at 90\% confidence level (C.L.) when muon and pion decay based neutrino beams are aimed at a 187~kton far detector and 1~kton near detector, both using the opaque detector technique. The near detectors are placed at 250~m from the sources. For comparison, the yellow area shows the excluded regions at 90\% CL$_s$ for the Daya Bay and Bugey-3 reactor neutrino experiments~\cite{Adamson:2020jvo}, whereas the red curves show the expected sensitivities for nuSTORM at 90\% C.L. The regions allowed by the gallium and reactor anomalies at 2$\,\sigma$ C.L.~\cite{Mention:2011rk,Giunti:2010zu} are shown by the blue and light blue areas, respectively.}
\end{figure*}

We conclude this subsection with a brief remark on the probable size of an opaque detector. As one can see from Figures~\ref{CPVplot_e-eff} and~\ref{CPVplot_sbratio}, the suitable detector size to study {\sl CP} violation depends on both the detection efficiency for electron-like events and the estimated signal-to-background ratio. In order to reach 3$\,\sigma$ C.L. sensitivity for at least 75.6\% of $\delta_{CP}$ values, the fiducial mass of the opaque detector needs to be about 120~kton and 130~kton for pion and muon beams, when the optimistic value of 70\% efficiency is assumed for electron-like events. For a more conservative assumption of 50\% efficiency, the required detector masses would be about 190 kton and 220 kton, respectively. The desirable detector mass is therefore about 120--220~kton, which may be feasible for an opaque detector.

\subsection{Sensitivity to active-sterile neutrino mixing}
\label{Sec:4.2}

In the so-called short-baseline anomaly, a significant departure from the standard three-neutrino oscillation picture have been observed. The reported excess of electron-like events reported in the experiments can be explained with the oscillations of three active neutrinos and one sterile neutrino corresponding to the mass-squared difference $\Delta m_{41}^2 \simeq$ 1~eV$^2$. Similar anomalies have also been noted in gallium and reactor experiments. Although the results have been revisited in numerous accounts, none has been able to confirm or refute the sterile neutrino hypothesis. It is left to the future neutrino oscillation experiments to check whether the observed excess indeed arises from active-sterine neutrino oscillations.

In this subsection, we study the effect of the selected detector parameters in the light sterile neutrino sensitivities. We calculate the expected sensitivity of the proposed nuSTORM experiment to provide a comparison and sensitivity target for our simulation study. We finally compare these sensitivities to determine the minimum requirements for the detector parameters. We examine the sensitivity to the 3+1--neutrino model in muon- and pion-beam experiments, while assuming a 187~kton opaque detector as the far detector. To optimize the experimental setup to look for active-sterile neutrino oscillations of $\Delta m_{41}^2 \simeq$ 1~eV$^2$, a near detector of variable mass is assumed. In our simulation, the near detectors are placed at 250~m distance from the J-PARC and MOMENT facilities, respectively. The chosen baseline length corresponds approximately to the conditions where the first oscillation maximum occurs in the $\nu_\mu \rightarrow \nu_s$ channel for $\Delta m_{41}^2 \simeq$ 1~eV$^2$ in J-PARC and MOMENT configurations, where $\nu_s$ is the sterile neutrino.

\begin{figure*}[t]
\begin{center}
\includegraphics[width=\linewidth]{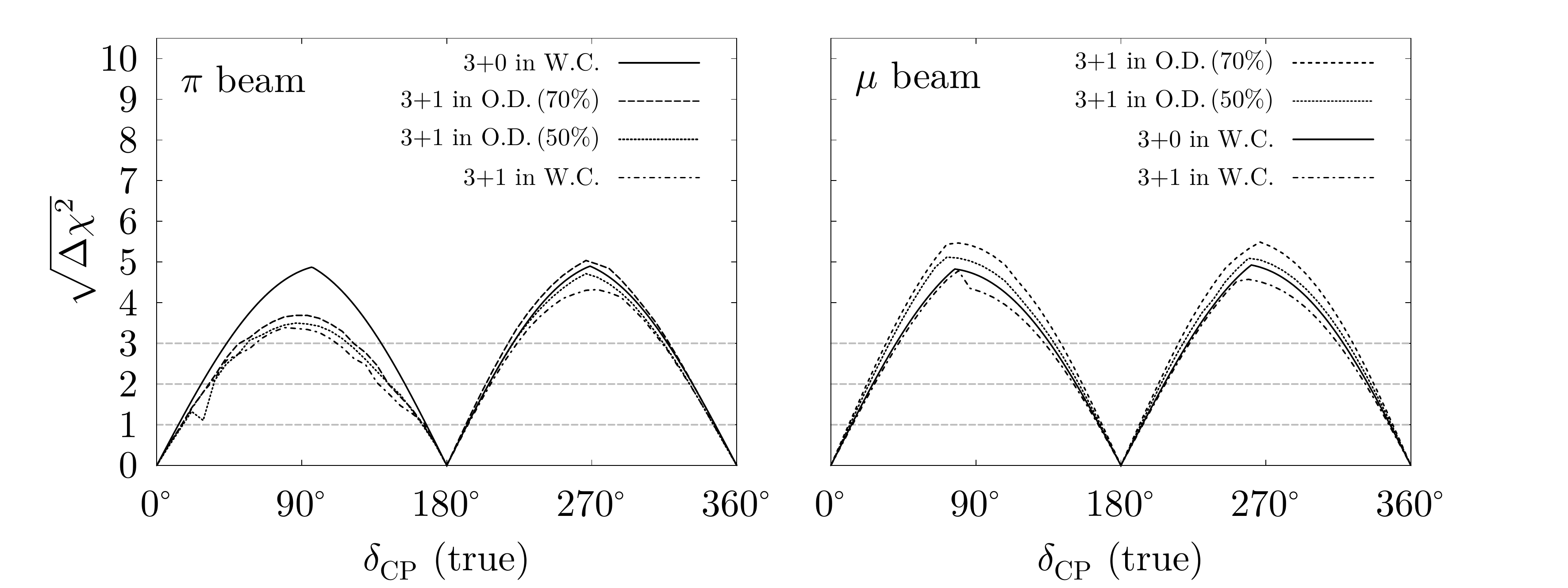}
\end{center}
\caption{\label{CPVplot_sterile} The {\sl CP} violation discovery potential in presence of a sterile neutrino. Both muon- and pion-decay based neutrino beams used with a 187~kton far detector based on water Cherenkov (W.C.) and opaque detector (O.D.) techniques. The sensitivities are provided both in 3+0 and 3+1--neutrino models. The opaque detector is considered for 50\% and 70\% efficiencies for $e$-like events. The sterile neutrino corresponds to the true values $\theta_{14} =$ 7$^\circ$, $\theta_{24} =$ 9$^\circ$ and $\Delta m_{41}^2 =$ 1~eV$^2$}.
\end{figure*}

We compare the opaque detector performance with the present experimental constraints on the sterile neutrino mixing as well as with the expected sensitivities of the nuSTORM experiment~\cite{Tunnell:2012nu,Adey:2014rfv}. NuSTORM is a next-generation short-baseline muon-decay-based neutrino oscillation experiment which has been proposed to search for signatures of sterile neutrinos in the LSND-MiniBooNE anomaly region. The experiment design includes a beamline with 100~kW power and a far detector that is placed at 2~km from the source. The beamline creates electron neutrinos and muon antineutrinos via the $\mu^+$ decay. The 1.3~kton far detector is based on the magnetized iron technology. We simulate the nuSTORM experiment by taking the neutrino fluxes and cross sections from Ref.~\cite{Tunnell:2012nu}, whereas the efficiencies and systematic uncertainties are adopted from Ref.~\cite{Adey:2014rfv}. The energy resolution function is taken to be $\sigma_e = 0.15\,E$. The experiment is assumed to run in $\mu^+$ mode for five years.

Figure~\ref{SterilePlot} presents the exclusion limits on $\Delta m_{41}^2$ and $\sin^2 2\theta_{14}$ as well as for $\sin^2 2\theta_{14}$ and $\sin^2 \theta_{24}$, where $\theta_{14}$ and $\theta_{24}$ are mixing angles characterizing the mixing active and sterile states. The $\chi^2$ fitting has been done by marginalizing the $\chi^2$ function over all parameters except for the ones shown in the axis labels.  In the left panel of the figure, the parameter values to the right of the contours are excluded for $\Delta m_{41}^2$ and $\sin^2 2\theta_{14}$ by 90\% C.L. if the MOMENT and J-PARC beams are used together with opaque detectors. The lower part of the curve is due to the 187~kton far detector, while the upper part arises from the presence of the 1~kton near detector. For comparison, the parameter values allowed by the gallium and reactor anomalies within 2$\,\sigma$ C.L.~\cite{Giunti:2010zu,Mention:2011rk} are shown with the blue and light blue areas. We also plotted the 90\% CL$_s$ limits obtained from the Daya Bay and Bugey-3 experiments~\cite{Adamson:2020jvo}. The excluded region is indicated with the yellow color. In the right panel of Figure~\ref{SterilePlot}, we show the exclusion contours at 90\% C.L. for $\sin^2 2\theta_{14}$ and $\sin^2 2\theta_{24}$, while $\Delta m_{41}^2 \sim$ 1~eV$^2$. In this region sensitivity is acquired almost entirely from the near detector. In both panels, we provide the sensitivities for the nuSTORM experiment, for which the exclusion limit is shown for 90\% C.L. with the solid red curve. While the confidence levels are calculated with the frequentist method, CL$_s$ statistics is defined in~\cite{Adamson:2020jvo}.

It is observed from Figure~\ref{SterilePlot} that both in $\Delta m_{41}^2-\sin^2 2\theta_{14}$ and $\sin^2 2\theta_{14}-\sin^2 2\theta_{24}$ panels the MOMENT beam proves to be more sensitive to the sterile neutrino parameters than the beam simulated for J-PARC. This behavior is in contrast to the results we obtained for the {\sl CP} violation discovery, where J-PARC is superior. Both beam configurations are sufficient to reach the nuSTORM sensitivities and to exclude large portions of the parameter values that are currently allowed by the gallium and reactor anomalies at 2$\,\sigma$ confidence level. We note that the projected sensitivities from nuSTORM can be reached on the $\sin^2 2\theta_{14}-\Delta m_{41}^2$ plane with the J-PARC and MOMENT configurations where an opaque detector of about 20~tonnes is used as the near detector. In a similar manner, we observe that the parameter values currently allowed by the gallium anomaly within 2$\,\sigma$ C.L. can be excluded by at least 90\% C.L. with the fiducial mass of about 10~tonnes. The reactor anomaly on the other hand cannot be completely excluded. The results are obtained for $\Delta m_{41}^2 \simeq$ 1~eV$^2$.

The upper limits on $\sin^2 2\theta_{14}$ and $\sin^2 \theta_{24}$ have recently been analysed using data from the Daya Bay, Bugey-3, MINOS and MINOS+ experiments~\cite{Adamson:2020jvo}. The Daya Bay data used in the analysis is collected from 1230 days of data taking. The report sets the current bounds to $\sin^2 2\theta_{14} \lesssim$ 5$\times$10$^{-2}$ and $\sin^2 \theta_{24} \lesssim$ 6$\times$10$^{-3}$ for $\Delta m_{41}^2 \simeq$ 1~eV$^2$ at 90\% CL$_s$, placing the most stringent limits on the active-sterile neutrino parameters. Our results in Figure~\ref{SterilePlot} show that a similar sensitivity can be reached using a 1~kton opaque detector in the J-PARC setup as a near detector, whereas MOMENT could provide higher sensitivities. We can therefore conclude that a near detector of at least 1~kton fiducial mass could improve the present bound on $\sin^2 2\theta_{14}$, whereas the current bound on $\sin^2 \theta_{24}$, which corresponds to $\sin^2 2\theta_{24} \lesssim$ 2.4$\times$10$^{-2}$ at 90\% C.L., is beyond reach.

We also studied the effect of changing the signal-to-background ratio and energy resolution in the results presented in Figure~\ref{SterilePlot}. We conclude that these parameters have negligible effects on the exclusion limits in the studied configuration.

The presence of the light sterile neutrino can undermine accelerator neutrino experiments in their mission to discover {\sl CP} violation in the leptonic sector. It has been shown that the inclusion of a light sterile neutrino can significantly hinder the discovery potential to {\sl CP} violation. In the following, we study its effect on the experimental configurations considered in this work.

We investigated whether the light sterile neutrino of $\Delta m_{41}^2 \sim \mathcal{O}$(1~eV$^2$) can induce significant loss in the sensitivity to {\sl CP} violation discovery, and if opaque detectors could help to recover the sensitivity. To do this, we calculated the {\sl CP} violation discovery potential as a function of $\delta_{CP}$ values both in the three-neutrino paradigm and in the 3+1 scenario, where a sterile neutrino corresponding to $\theta_{14} =$ 7$^\circ$, $\theta_{24} =$ 9$^\circ$, $\theta_{34} =$ 0$^\circ$ and $\Delta m_{41}^2 =$ 1~eV$^2$ is assumed in the model. The $\chi^2$ fitting is done by marginalizing over all parameters except for $\delta_{CP}$, $\theta_{34}$, $\delta_{24}$ and $\delta_{34}$. The results are presented in Figure~\ref{CPVplot_sterile}, where the sensitivities to the {\sl CP} violation are shown for both water Cherenkov and opaque detector setups.

As can be seen in Figure~\ref{CPVplot_sterile}, the presence of the active-sterile oscillation parameters $\theta_{14}$, $\theta_{24}$ and $\Delta m_{41}^2$ causes a remarkable drop in the {\sl CP} violation discovery potential. When a 187~kton far detector based on water Cherenkov technique is assumed without a near detector, the sensitivity at $\delta_{CP} =$ 270$^\circ$ drops from 4.9$\,\sigma$ to 4.3$\,\sigma$ confidence level for J-PARC and from 4.8$\,\sigma$ to 4.5$\,\sigma$ for MOMENT, respectively. Replacing the water Cherenkov vessel with an opaque detector of 70\% efficiency and equivalent mass improves the sensitivity at $\delta_{CP} =$ 270$^\circ$ from the water Cherenkov limits to 5.0$\,\sigma$ and 5.4$\,\sigma$ C.L. in the 3+1 model, restoring the lost sensitivity and even improving it beyond the 3+0 model limits in both J-PARC and MOMENT setups.

\section{Conclusions}
\label{Sec:5}
We investigated the prospects and requirements of using an opaque detector in accelerator neutrino experiments, outlining its potential in neutrino beams produced via pion and muon decays. Using the future J-PARC and MOMENT beam facilities as examples of both neutrino production techniques, we identified the detector properties where the opaque detector technology could provide an advantage. We then examined whether or not these properties have effects on the {\sl CP} violation and sterile neutrino sensitivities. We discovered that the opaque detector could enhance the physics reach in such experiments through its improved particle identification capabilities, but not through its low energy threshold.

After identifying the strengths of the opaque detector technology, we determined the minimum requirements for the J-PARC and MOMENT beam experiments to reach their target sensitivities with an opaque detector. For the {\sl CP} violation search, we compared the sensitivities with the future T2HK experiment, while the sterile neutrino sensitivities were compared with the recently proposed nuSTORM experiment. We found that for an opaque detector of 187~kton fiducial mass, the minimum requirements to reach the {\sl CP} violation sensitivities for the electronlike efficiency are 46\% and 51\% for the J-PARC and MOMENT configurations, whereas the requirements for the signal-to-background ratios are 20 and 5, respectively. In our baseline simulation of MOMENT and J-PARC, we assumed 50\% efficiencies for electron-like events. The estimated signal-to-background ratios with the 187~kton far detectors are about 20 and 7 with the opaque detector technology, where the simulations assuming the standard water Cherenkov detector involved ratios of 4 and 5, respectively. We also remark that a near detector of about 20~tonnes fiducial mass and 250~m baseline is adequate to reach the nuSTORM sensitivities and exclude the parameter values allowed by  2$\,\sigma$ C.L. in the gallium anomaly.

Finally, we studied whether an opaque detector could also improve the sensitivity to {\sl CP} violation in presence of the sterile neutrino. We found that using a 1~kton near detector and 187~kton far detector improves the sensitivities from the corresponding water Cherenkov setup. When the opaque detectors are used, the statistical significance increases from 4.3$\,\sigma$ to 5.0$\,\sigma$ confidence level for J-PARC and from 4.5$\,\sigma$ to 5.4$\,\sigma$ for MOMENT, respectively, when $\delta_{CP} =$ 270$^\circ$ and the 3+1 model are assumed.

Altogether, we have found the opaque detector an attractive detector candidate for accelerator-based neutrino experiments. Although neutrino energies above 100~MeV and below 1~GeV are a very challenging regime to the LiquidO technique, opaque detectors could offer a way to reach {\sl CP} violation sensitivities much higher than those of water Cherenkov detectors, if an adequate interplay of detector mass, efficiency and signal-to-background ratio is achieved. Such a configuration could become feasible by using Gd-doping in opaque scintillator, which is known to be much easier to load than water. One could also consider using other detector technologies for these experiments. However, the scalability and adaptability make the opaque scintillator a competitive choice. The far detectors in accelerator neutrino experiments ask for a large size and long durability, which give opaque scintillators an advantage over transparent scintillators. A dedicated study on the detector response is needed for
the opaque detectors, though. Our conclusions should encourage the opaque detector working groups to press forward with the development of the technique.

\section*{Acknowledgment}
This work was supported in part by Guangdong Basic and Applied Basic Research Foundation under Grant No. 2019A1515012216, National Natural Science Foundation of China under Grant Nos. 11505301 and 11881240247, China Postdoctoral Science Foundation under Grant No. 2020M672930, the university funding based on National SuperComputer Center-Guangzhou (74130-31143408) and the CAS Center for Excellence in Particle Physics (CCEPP). We thank Dr. Anatael Cabrera for his insightful comments on the first version of our manuscript. The MOMENT accelerator working group is highly appreciated for providing the flux files for the MOMENT beam.

\bibliography{bibliography}

\end{document}